\documentclass[aps,prstab,reprint,groupedaddress,showpacs,showkeys]{revtex4-1}

\usepackage{wrapfig}
\let\mult=\cdot
\usepackage[squaren,Gray]{SIunits}
\let\cdot=\mult
\usepackage{textcomp}
\usepackage[percent]{overpic}
\usepackage{overpic}
\usepackage[center]{subfigure}
\usepackage{graphicx}
\usepackage{psfrag}
\usepackage{url}
\usepackage{hyperref}
\usepackage{breakurl}
\usepackage{amssymb, amsmath} 
\usepackage{multirow}
\usepackage{tabularx}

\newcommand{\sub}[1]{\ensuremath{_{\mbox{\protect\scriptsize{#1}}}}}
\newcommand{\rs}{$R\sub{s}$ }
\newcommand{\rres}{$R\sub{res}$ }
\newcommand{\rbcs}{$R\sub{BCS}$ }
\newcommand{\tc}{$T\sub{c}$ }		

\begin{document}

\title{High-Q operation of SRF cavities: The potential impact of thermocurrents on the RF surface resistance}
\author{J.-M. Vogt}
\email{julia.vogt@helmholtz-berlin.de}
\author{O. Kugeler}
\author{J. Knobloch}
\affiliation{Helmholtz-Zentrum Berlin, Germany}
\author{(submitted to PRST-AB 9 October 2014)}

\begin{abstract}
For many new accelerator applications, superconducting radio frequency (SRF) systems are the enabling technology. In particular for CW applications, much effort is being expended to minimize the power dissipation (surface resistance) of niobium cavities. Starting in 2009, we suggested a means of reducing the residual resistance by performing a thermal cycle \cite{Kugeler2009}, a procedure of warming up a cavity after initial cooldown to about $20\,\kelvin$ and cooling it down again. In subsequent studies \cite{Vogt2013a}, this technique was used to manipulate the residual resistance by more than a factor of $2$. It was postulated that thermocurrents during cooldown generate additional trapped magnetic flux that impacts the cavity quality factor. Here, we present a more extensive study that includes measurements of two additional passband modes and that confirms the effect. In this paper, we also discuss simulations that support the claim. While the layout of the cavity LHe tank system is cylindrically symmetric, we show that the temperature dependence of the material parameters results in a non-symmetric current distribution. Hence a significant amount of magnetic flux can be generated at the RF surface.
\end{abstract}

\maketitle

\section{Introduction\label{Chap:Intro}}
Much SRF research and development is presently concentrating on finding means to reduce the losses in the cavity for next-generation accelerator applications. It was found that the surface resistance \rs can be impacted by a number of different treatments. \rs consists of a temperature dependent BCS term \rbcs and a residual resistance $R\sub{res}$, both of which may be impacted by the treatment. For example, it was discovered that the annealing of high RRR niobium with titanium or nitrogen can significantly change \rs and especially the dependence of \rs on the RF magnetic field \cite{Grassellino2013,Dhakal2013}.

We expand here on studies \cite{Vogt2013a,Kugeler2009,Kugeler2013,Eichhorn2013, Gonnella2015} to minimize the additional residual term in the cavity systems introduced by thermal currents based on thermal cycling. During a thermal cycle the superconducting cavity is warmed up to temperatures around $20\,\kelvin$ which exceeds the transition temperature $T\sub{c}$. Afterwards the cavity is cooled down again to the operating temperature. Depending on the temperature \textit{difference} between the two ends of the cavity tank system during the superconducting (sc) phase transition, significant variations of the surface resistance are observed. In the best case the residual resistance of the $1/9\,\pi$ mode is decreased from $10.6\,\nano\ohm$ to $1.4\,\nano\ohm$ which is more than a factor of seven.

In contrast to annealing techniques, the cycling procedure should not change \rbcs because the material properties such as mean free path and coherence length are not influenced. Hence we attribute the observed modification of \rs for different cycles to a change in residual resistance. This assumption is borne out by extracting the residual resistance from quality factor versus temperature measurements (see Section \ref{sec:rres}).

\subsection{Thermoelectric effect in the cavity tank system}

The TESLA-type cavity reported on here received a heavy BCP (about $150\,\micro\meter$) prior to a $2\,\hour$ bakeout at $800^{\circ}$C.  A light BCP etch followed the heat treatment. Before the helium tank was welded onto the cavity a quality factor of about $2\cdot10^{10}$ in the $\pi$ mode at $2\,\kelvin$ was measured in a vertical test which corresponds to a residual resistance of $\approx\,1.2\,\nano\ohm$ if one assumes that \rbcs did not change between vertical and horizontal test (fitting parameters for \rbcs in horizontal test are listed in Section \ref{sec:rres}). In all tests presented in this paper, the cavity was installed horizontally in the HoBiCaT facility \cite{Kugeler2010}, similar to an installation in a cryomodule. 

In a previous study \cite{Vogt2013a}, we suggested that thermoelectrically induced currents and their associated magnetic flux is responsible for the change of \rs upon thermal cycling. It is dependent on the temperature difference along the cavity helium tank system during cooldown. In the horizontal setup, the system is fabricated of two materials: Niobium (cavity) and titanium (helium tank). If a temperature gradient is applied along the system, a voltage arises which drives a current along the cavity and back through the tank:
\begin{eqnarray}
V &=& (S\sub{Nb} - S\sub{Ti}) \cdot (T\sub{1} - T\sub{2}) \label{eq:seb} \\
&=& I \cdot R\sub{system, DC} 
\end{eqnarray}
where $S$ is the thermopower or Seebeck coefficient of the respective material, $T\sub{1,2}$ are the temperatures of the two joints, $I$ is the thermocurrent and $R\sub{system, DC}$ is the DC resistance of the system which will be dominated by the resistivity of the helium tank, especially once part of the cavity goes superconducting.

The magnetic flux associated with the current can then be trapped in the niobium when it becomes superconducting and hence produces additional RF losses \cite{Vallet1992,Aull2012,Gurevich2013}.

A number of questions were left open in our previous paper. For one, the thermopower for niobium and titanium were poorly known so reliable estimates of the thermocurrents near \tc were not possible. Furthermore, it was not clear how magnetic flux at the RF surface can be generated by a cavity tank system whose geometry is cylindrically symmetric. Finally, a systematic study of the surface resistance versus (controlled) temperature difference during the superconducting transition was still lacking.

Here we present a far more systematic and in-depth study that addresses these questions. In the first part (Section \ref{Chap:Cav}) we discuss more extensive measurements of the quality factor of a TESLA cavity for many temperature differences. Heaters were installed to generate temperature differences between the cavity ends that are typically encountered during cooldowns in HoBiCaT. These measurements indeed confirmed a direct correlation between the residual resistance and the temperature difference (see Section \ref{sec:correl}). We also demonstrated that a reduced temperature difference ($10\--20\,\kelvin$) between the cavity ends during the very first cooldown of the cavity to the sc state can produce high quality factors without an additional thermal cycle being required. These results were confirmed for several modes of the passband and suggest that Q factors up to $10^{11}$ are possible in our test cryostat, especially for the center cells where the external magnetic shield is most effective.

In the second part (Section \ref{sec:sims}) we discuss a quantitative analysis of the system to determine if indeed the thermoelectric effect can explain the observed cavity behavior. Here, we turned to thermoelectric simulations of the cavity with helium tank. For these, accurate values are required for the temperature dependent thermoelectric properties of both niobium and titanium. Measurements presented in Section \ref{sec:thermopower} now provide more reliable values of the thermopower that were used in simulations.

The simulations allow one to calculate the thermoelectric currents in the cavity tank system. Mechanically, the system exhibits toroidal symmetry (ignoring misalignment issues). Assuming perfect rotational symmetry one would then expect a magnetic field between cavity and tank in azimuthal direction. Its magnitude is zero at the inner cavity RF surface, rising inside the wall, having a maximum at the outer surface of the cavity and falling back to zero on the outer surface of the tank. Hence it should not influence the RF properties. However, what counts is the symmetry of the current flowing in the system which in fact can be broken due to the temperature dependence of the material properties, as discussed in Section \ref{sec:sym}. Due to the broken symmetry, magnetic flux actually exists at the RF surface and within the cavity volume that potentially can be trapped. Our simulations, whose results are presented in Section \ref{sec:simres}, confirm that the expected thermocurrent distribution is consistent with the observed variations in cavity quality factor.

\subsection{Trapping of ambient magnetic field} \label{sec:trapp}

While we focus on the impact of thermocurrents and explain why it is a plausible explanation for the observed variation of \rs in our experiments, there is an additional effect that was intensively investigated during the last months: The change of the efficacy of the Meissner effect depending on the cooling rate and temperature distribution for slow, homogeneous cooling \cite{Aull2012,Romanenko2014c,Romanenko2014b,Gonnella2014,Gonnella2015}. This effect can therefore also play a role in determining the residual resistance in cavities.

Since our instrumentation is limited to sensors outside the helium tank we cannot make any statement on the exact temperature \textit{distribution} of the cavity. However, we believe that a change in flux trapping efficiency of the residual field has only a minor impact on our particular results. Our cavity was shielded from the earth's magnetic field by a double (warm/cold) shield combination, and the ambient field at the cavity position, as measured at room temperature, is low (typically a fair amount less than $0.5\,\micro\tesla$ at the center cells, about $1\,\micro\tesla$ maximum in the end cells \cite{Kugeler2010}). The inner cold shield was manufactured from Cryoperm, so that the shielding efficacy should theoretically be even better during cold operation.  We measured the permeability of the material \cite{Kugeler2011} and actually found little temperature dependence, so that the warm results should in fact be a reasonable estimate of the residual field during cavity operation.  This is borne out by the best residual surface resistances of only $1.4\,\nano\ohm$ observed in the center cells of the cavity when the temperature difference during cooldown is small and the thermoelectric effect should play no role. 

Furthermore we stress that the cavity we tested was not nitrogen doped. While doped as well as undoped cavities trap a higher percentage of ambient field when the cooling rate is decreased below a certain value, the same amount of trapped flux causes considerably less RF losses in an undoped cavity \cite{Gonnella2014}. Hence the difference between trapping all of the ambient flux in the center cells (at most $0.5\,\micro\tesla$) and not trapping any of the flux would lead to a decrease of at most $1.75\,\nano\ohm$ in \rs for the $1/9\,\pi$ mode using the approximation of $3.5\,\nano\ohm / \,\micro\tesla$ \cite{Aune2000}. This is considerably less than the observed change of $9.2\,\nano\ohm$ between thermal cycles. For the $\pi$ mode the contribution may be larger and this is one of the reasons we measured several modes of the passband.  We show later (e.g., Figure \ref{fig:ModeCompare}) that indeed we observe no significant impact of a changing Meissner efficacy on our measurement results.

\begin{figure}
\includegraphics[width=\linewidth]{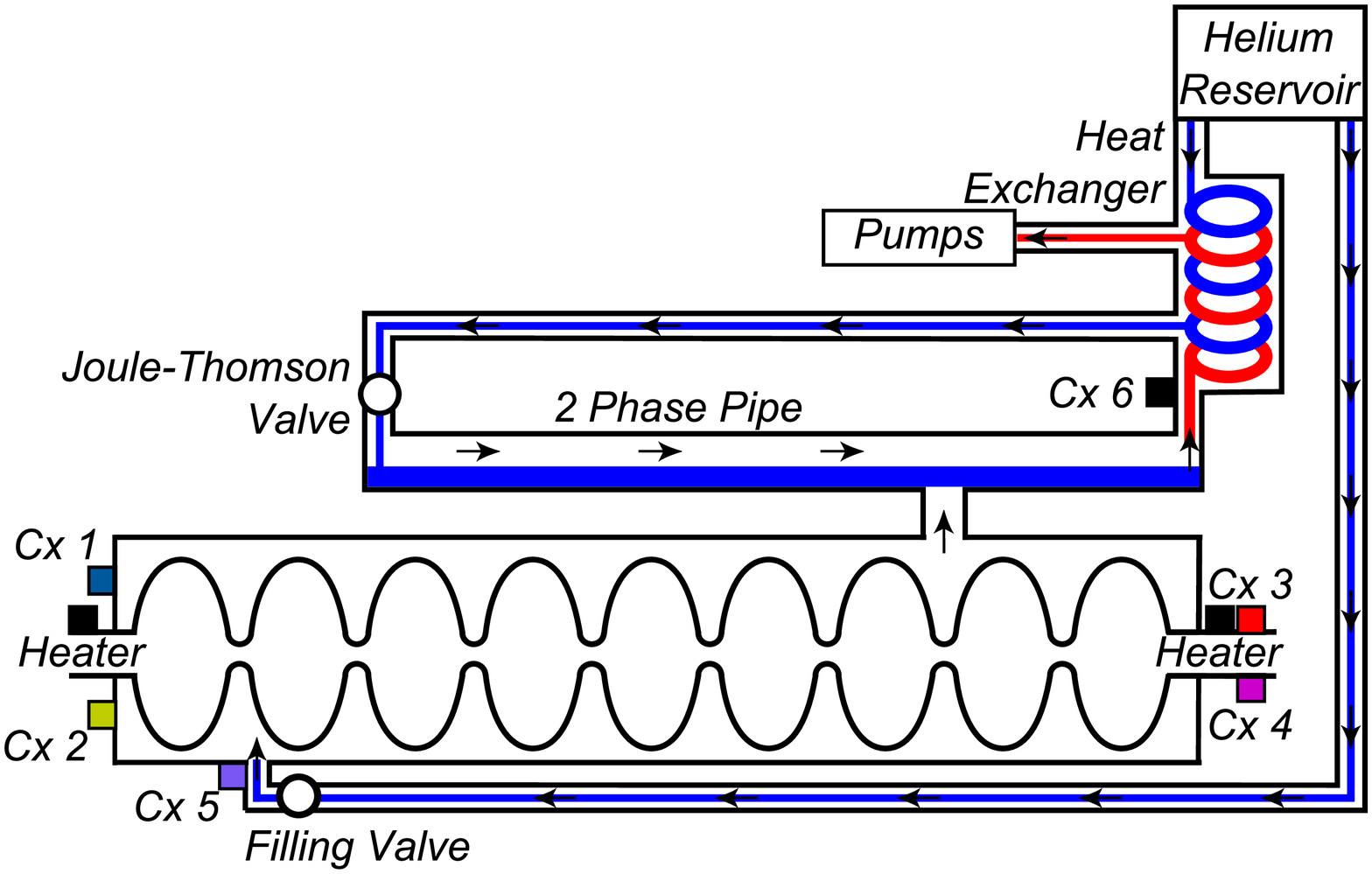}
\caption{Setup of TESLA cavity inside HoBiCaT cryostat with positions of Cernox temperature sensors 1 -- 6 and heaters.\label{fig:setup}}
\end{figure}

\section{Thermal cycling experiments\label{Chap:Cav}}

\subsection{Experimental setup and cooling procedures} 

A fully equipped TESLA-type cavity welded into a titanium tank and with a TTF-3 input coupler installed was mounted horizontally inside the HoBiCaT cryostat. The cavity was equipped with Cernox sensors on the helium vessel head and beam pipes near the Nb-Ti joints, as well as helium inlet and outlet. Furthermore two heaters were attached, one on each beam pipe. The setup including the helium supply is sketched in Figure \ref{fig:setup}.

HoBiCaT can cool the cavity with different schemes. The cryoplant fills the helium tank via two valves: The filling valve (FV) at the bottom left and the Joule-Thomson valve (JTV) which fills the tank via the 2-phase-pipe from the top right. We used three different cooling schemes, the initial cooldown, the thermal cycle and the parked cooldown, which will be explained in the next sections.

\begin{figure}
\includegraphics[width=\linewidth]{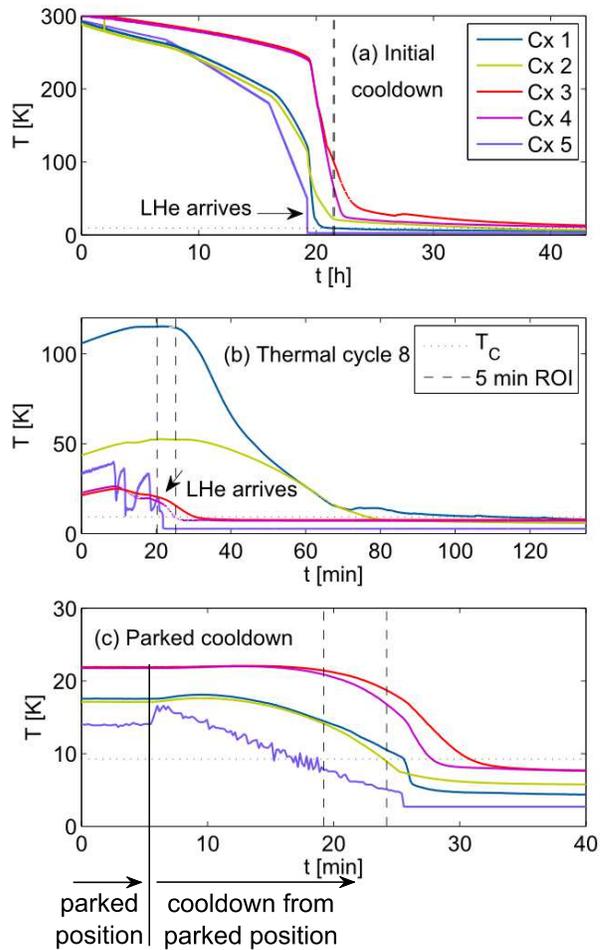}
\caption{Temperatures measured by Cx 1 -- Cx 5 during initial cooldown (a), a thermal cycle (b) and parked cooldown (c). The dotted line indicates the transition temperature of $9.2\,\kelvin$. The dashed line indicates the time when the first sensor dropped below transition temperature and a $5\,\minute$ interval before that transition. \label{fig:Ts}}
\end{figure}

During the \textbf{initial cooldown} (whose temperature profile is shown in Figure \ref{fig:Ts}a), mainly the FV at the bottom of the cavity was used for filling the helium tank with $4.2\,\kelvin$ helium. The coldest spot was always the one closest to the FV. This lead to high spatial temperature gradients along the cavity \textit{and} from bottom to top. The JTV was $25\,\%$ open as well to cool down the heat exchanger but no liquid helium reached the cavity. 

For a subsequent \textbf{thermal cycle} (as in Figure \ref{fig:Ts}b) the FV was closed and remained so during the whole process. As the first step the JT valve was closed and the heaters on both ends were used to evaporate the helium in the tank and create a temperature difference between the cavity ends if desired. The targeted difference could be adjusted by varying the heater power. Values chosen were typical of those encountered during normal cooldowns. Cernox sensor 5 (purple) measured a jump in temperature once the tank was empty and the cavity was normal conducting. In addition, we monitored the temperature of the evaporated gas with Cernox sensor 6 (black) which exhibited a characteristic rise in temperature (sensor is not displayed in Figure \ref{fig:Ts} for clearness). Now, the JT valve was opened again to start gas flow through the system and to restart cooling while the FV remained closed. 

To pinpoint the time of the phase transition we measured the bandwidth of the $\pi$ mode during one complete warm-up/cooldown cycle.  This allowed us to correlate the sc transition with the temperatures shown by Cernox sensors. The first sensor at the ends read close to $9.2\,\kelvin$ within less than one minute of the time when the first cells of the cavity went superconducting.

Note that cycling scheme may lead to a reversed spatial temperature difference compared to the initial cooldown if not balanced by the heaters because the cold gas and liquid started to pour from the top right. As soon as the first liquid entered the tank a drop in temperature was registered by Cernox sensor 5 as can be seen in Figure \ref{fig:Ts}.

During one test series we performed a \textbf{``parked cooldown''} from room temperature which combines properties of both the initial cooldown and the thermal cycle (Figure \ref{fig:Ts}c). The cooling procedure of the initial cooldown was adapted to stop well before the sc phase transition. The cryoplant was balanced to maintain a constant temperature at the FV for $48\,\hour$. The set point was first set to $30\,\kelvin$ and then continuously lowered to $14\,\kelvin$ during this period. Both valves, the FV and the JTV, were used. After all temperature sensors were clearly in equilibrium (no temperature change with time) the set point was further lowered towards $1.8\,\kelvin$ and the cavity tank system transitioned with a small $\Delta\,T\,<\,10\,\kelvin$ into the sc state.

Altogether, we performed one initial cooldown from room temperature, followed by 11 cycles, one parked cooldown from room temperature and another two cycles. After each cycle we performed a measurement of quality factor $Q$ versus temperature at a ($\pi$ mode) gradient of $4\,\mega\volt/\meter$ to extract the residual resistance for all three modes.

The coupling during the RF cold tests was in most cases close to critical and always between $0.3$ and $3.5$ which results in a low error margin below $10\,\%$ \cite{Powers2005}. A three stub tuner was installed in front of the input antenna and used to ensure optimal coupling for each mode.

\begin{figure}
\includegraphics[width=1.0\linewidth]{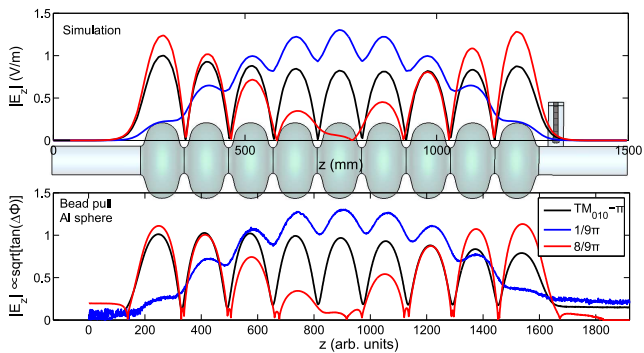}
\caption{Field distributions of three passband modes of a TESLA cavity: $\pi$ mode, $1/9\,\pi$ mode and $8/9\,\pi$ mode computed with CST and measured with a bead pull setup using an Al sphere.\label{fig:Fields}}
\end{figure}

\subsection{Cavity modes}
For the evaluation of the residual resistance in the various cells of the cavity we measured the quality factor in three different passband modes: The $\pi$ mode ($1299\,\mega\hertz$), the $8/9\,\pi$ mode (1298\,\mega\hertz) and the $1/9\,\pi$ mode (1274\,\mega\hertz). Each measurement is therefore an average over the cells which are exposed to the RF field in the respective mode (${R\sub{s}}\,=\,G/Q\sub{0}$). The geometry factor is taken from CST simulations:
\begin{eqnarray*}
 G_{\pi}&=&271.2\,\ohm\\
 G_{8\pi/9}&=&271.5\,\ohm\\
G_{\pi/9}&=&268.3\,\ohm.
\end{eqnarray*}
The $\pi$ mode exhibits a homogeneous field distribution along the cavity. The $1/9\,\pi$ mode has the maximum field in the center cells and low field in both end cells whereas the $8/9\,\pi$ mode has minimum field in the mid cell and maximum field in the end cells.

Figure \ref{fig:Fields} shows a calculation of the field distribution in the cavity computed with CST microwave studio. The plot is normalized to the same stored energy. We validated the computed field distribution with a bead pull measurement after all RF measurements were completed (Figure \ref{fig:Fields}).

The bead pull measurement confirmed the overall field distribution of the three passband modes. Hence, they were used to evaluate contributions to the surface resistance depending on the location. We were able to distinguish between end cell region (dominant in $8/9\,\pi$) and mid cell region (dominant in $1/9\,\pi$) but we could not determine the cell number because all modes are mirror symmetric with respect to the cavity center plane in the axial direction. 

The RF measurements confirmed that \rbcs did not vary between the three modes (see Section \ref{sec:rres}). In contrast, the measured \rres was influenced by localized contributions like the spatial variation of HoBiCaTs ambient magnetic flux due to inhomogeneous shielding at the end cells and maybe material defects. A higher surface resistance in the end cells lead to a lower $Q$ in the $8/9\,\pi$ and $\pi$ mode than in the $1/9\,\pi$ mode. 

\begin{figure}
\includegraphics[width=\linewidth]{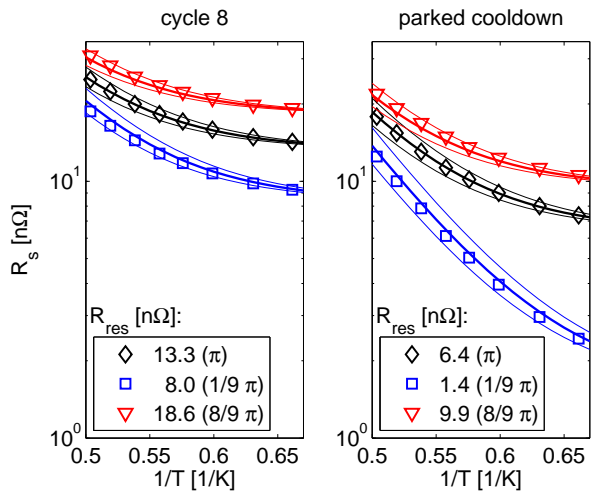}
\caption{Arrhenius plots of cycle 8 (highest $\Delta T$) and the parked cooldown (lowest $\Delta T$). The thick lines indicate the BCS fits with $A\,=\,31.8\,\micro\ohm$ and $B\,=\,15.7\,\kelvin$ for the stated $R\sub{res}$. The thin lines indicate the range of fits that are covered by the uncertainty of $A$ and $B$. \rs was calculated using $R\sub{s}\,=\,G/Q\sub{0}$ and $G_{\pi}\,=\,271.2\,\ohm$, $G_{8\pi/9}\,=\,271.5\,\ohm$, $G_{\pi/9}\,=\,268.3\ohm$.
\label{fig:BCS_example}}
\end{figure}

\subsection{Extraction of residual resistance} \label{sec:rres}
The temperature dependent BCS contribution to the surface resistance \rbcs was separated from the temperature independent residual contribution \rres using \cite{Ciovati2014}:
\begin{eqnarray}
R\sub{s}(T) &=& A \cdot \exp\left(\frac{-B}{T}\right) + R\sub{res}\\
\mbox{with}\quad B &=& \frac{U}{k\sub{B}}
\end{eqnarray}
For all measured modes, the fit parameters $A$ and $B$ overlapped within the error margin confirming that \rbcs remained constant for different cycles. Hence we determined overall BCS-parameters of $A\,=\,(31.8\,\pm\,2.2)\,\micro\ohm$ and $B\,=\,(15.7\,\pm\,0.2)\,\kelvin$ corresponding to $U\,=\,(1.70\pm0.02)\,k\sub{B} \tc\,=\,(1.35\pm0.02)\,\milli\electronvolt$. With these parameters we extracted the residual resistance from all measurements. Figure \ref{fig:BCS_example} shows two examples for BCS fits and \rres extraction: Cycle 8 (highest temperature difference during phase transition) and the parked cooldown (lowest temperature difference during phase transition). The error bars of the RF measurement are within the symbol size.

\subsection{Correlation of \rres with cooling parameters} \label{sec:correl}

Table \ref{tab:Qs} provides examples of the measured residual resistances. We observed that cooldowns with a low temperature difference along the cavity result in lowest residual resistances. For the $1/9\,\pi$ mode (field dominant in center cells) this amounted to a residual resistance of $1.4\,\nano\ohm$. Presumably the quality factor then is dominated by the ambient field in HoBiCaT which is about $0.2 - 0.5\,\micro\tesla$ along the center cells causing a residual resistance of $0.7 - 1.75\,\nano\ohm$ \cite{Aune2000}. Other, non-flux contributions may also add to the residual resistance.

\begin{figure}
\includegraphics[width=\linewidth]{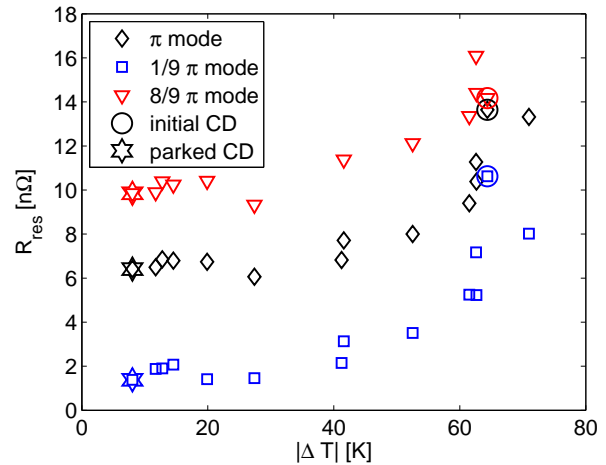}
\caption{Residual resistance measured with the three passband modes as a function of the temperature difference along the cavity at the onset of sc phase transition. The lowest residual resistance was achieved in the center cells following the cycle 1 and the parked cooldown ($1.4\,\nano\ohm\,\widehat{=}\,Q\sub{0}> 10^{11}$ at $1.5\,\kelvin$).\label{fig:RvsDT}}
\end{figure}

\begin{table}
\setlength{\tabcolsep}{2.8mm}
\renewcommand{\arraystretch}{1.5}
\begin{center}
\begin{tabularx}{\columnwidth}{m{2.5cm}ccc}
\hline
$R\sub{BCS}\,\approx\,0.9\,\nano\ohm$ & \multicolumn{3}{c}{$R\sub{res}$}\\
at $1.5\,\kelvin$ & $\pi$ mode & $8/9\,\pi$ mode & $1/9\,\pi$ mode\\
\hline
Initial cooldown & 13.6 & 14.2 & 10.6\\
\hline
Thermal cycle 8 (high $\Delta T$) &  13.3 & 18.6 & 8.0 \\
\hline
Parked cooldown & 6.4 & 9.9 & 1.4 \\
\hline
Thermal cycle 7 (low $\Delta T$)& 6.5 & 9.9 & 1.9\\
\hline
\end{tabularx}
\end{center}
\caption{Examples of measured residual resistances of the three passband modes.}
\label{tab:Qs}
\end{table}

Figure \ref{fig:RvsDT} displays the extracted residual resistances for all cooldowns and cycles as a function of temperature difference $\Delta T$. This is defined as:
\begin{eqnarray}
\Delta T\,=\,\left|\frac{T\sub{Cx1}+T\sub{Cx2}}{2}\,-\,\frac{T\sub{Cx3}+T\sub{Cx4}}{2}\right|
\end{eqnarray}
The temperature difference between the temperature sensors on the two ends of the tank is relevant for the thermocontact voltages. This parameter is a measure for the induced thermoelectric voltage and calculated at the instance when the first of the four sensors drops below the transition temperature.

Figure \ref{fig:RvsDT} shows that the residual resistance of the cavity decreases with $\Delta T$. The curves for the three passband modes run parallel indicating that all nine cells are similarly affected. Thus the change in \rres is a global effect consistent with the thermocurrent model. 

Furthermore, we see that the $1/9\,\pi$ mode has a significantly reduced surface resistance compared to the two other modes while the residual resistance of the $8/9\,\pi$ mode is elevated. 

We believe this is due to an increased ambient field in HoBiCaT near the end cells. Measurements of the magnetic shielding in HoBiCaT yielded that the magnetic field inside the end parts of the shield is increased due to cut outs for coupler and geometry effects (up to about $1\,\micro\tesla$) \cite{Kugeler2010}.

Note that we were limited to only four temperature sensors outside the tank which only provides an incomplete picture of the temperature distribution. Two sensors were near heaters which results in an additional offset in $\Delta T$. For more reliable values more sensors along the setup are needed.

\begin{figure}
\includegraphics[width=\linewidth]{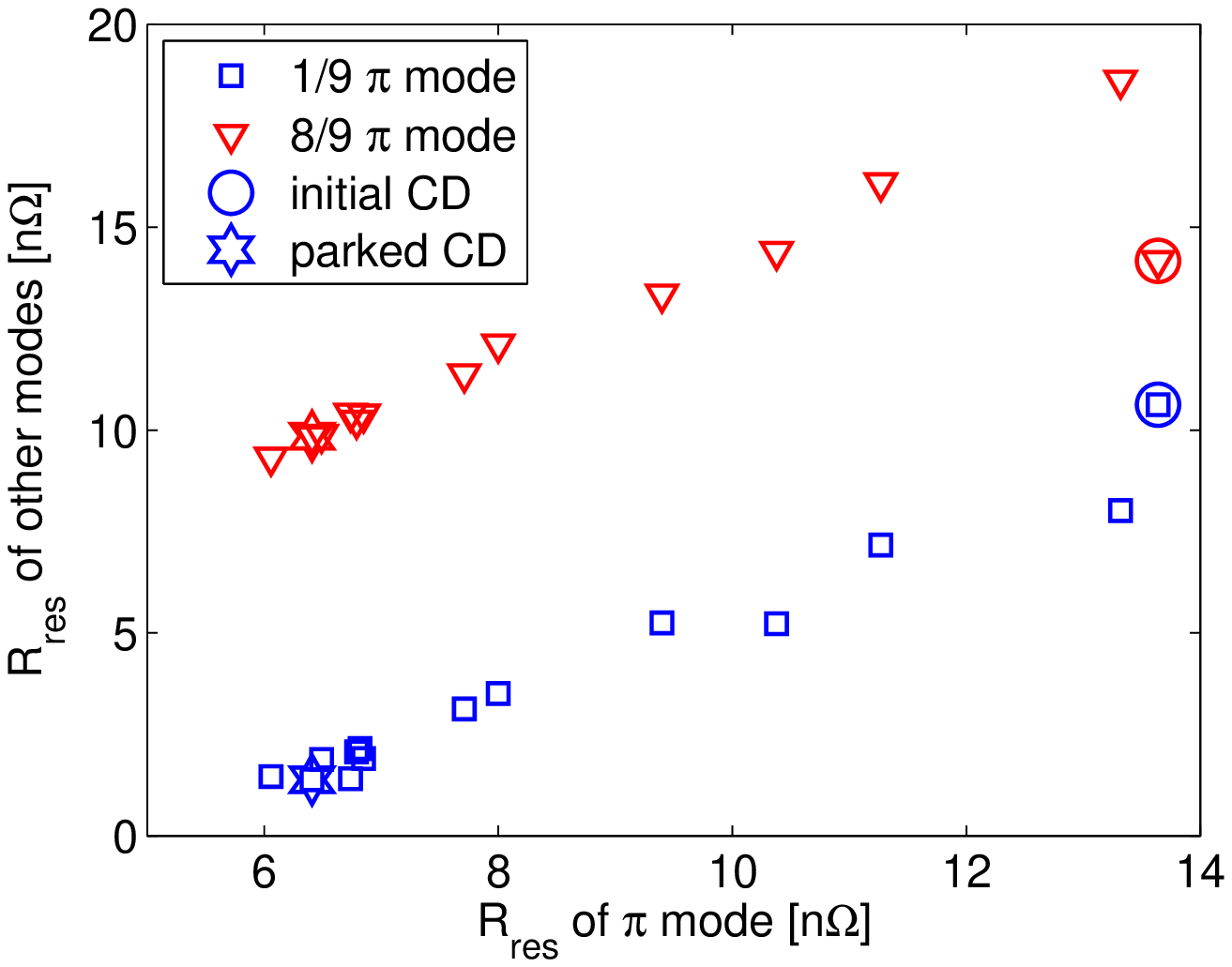}
\caption{\rres obtained for the $8/9\,\pi$ and $1/9\,\pi$ mode as a function of \rres obtained for the $\pi$ mode. Linear regression leaving out the initial cooldown results in: \\$R_{8/9\,\pi}\,=\,1.2\cdot R_{\pi} +1.8\,\nano\ohm$ and\\$R_{1/9\,\pi}\,=\,1.0\cdot R_{\pi} -4.7\,\nano\ohm$.\label{fig:ModeCompare}}
\end{figure}

Figure \ref{fig:ModeCompare} compares the average surface resistance of the modes to that of the $\pi$ mode. A linear dependency is visible for the cycles and the parked cooldown. The initial cooldown for the $8/9\,\pi$ does not fit into the linear curve which is already visible in Figure \ref{fig:RvsDT}. This might be explained by the fact that the cooling dynamics of initial cooldown are different from the other coolings. The effect would be largest for the cells in the higher ambient magnetic field region, consistent with the observation that the $8/9\,\pi$ mode differs significantly. 

Leaving out the data point of the initial cooldown, the two graphs in Figure \ref{fig:ModeCompare} exhibit a linear behavior. The slopes are $1.2$ ($8/9\,\pi$ mode) and $1.0$ ($1/9\,\pi$ mode) which is close to $1$ meaning that all cells are affected by the thermocurrent in a similar way as would be expected since the current is constant along the length of the cavity. The current and resulting trapped magnetic flux adds a constant residual resistance to each cell.

The results in Figure \ref{fig:RvsDT} and Figure \ref{fig:ModeCompare} also demonstrate that the efficacy of the Meissner effect in expelling the constant ambient field plays no major role in our observed changes in $R\sub{s}$.   Otherwise the $8/9\,\pi$ mode should change its residual resistance more than the $1/9\,\pi$ mode.  Then the three curves in Figure \ref{fig:RvsDT} would not run parallel and the slopes of the curves in Figure \ref{fig:ModeCompare} would not be linear with a slope near $1$. The exception is the $8/9\,\pi$ mode result after the initial cooldown ($\Delta R\sub{s}\,\approx\,5\,\nano\ohm$) which may be due to the fact that the cooling conditions during initial cooldown significantly differ from the conditions during the thermal cycles and the parked cooldown resulting in different Meissner efficacies. This would most affect the $Q$ value of the $8/9\,\pi$ mode since the measurement averages over the high-ambient-field end cells.

The intercepts in Figure \ref{fig:ModeCompare} are $+1.8\,\nano\ohm$ ($8/9\,\pi$ mode) and $-4.7\,\nano\ohm$ ($1/9\,\pi$ mode). They are a measure of the different residual resistance that is not caused by thermoelectrically induced magnetic flux but is always present. The values indicate that the end cells of the cavity have higher average surface resistance. Sources can be either the increased ambient magnetic field near the end cells in HoBiCaT as discussed above or particulate contamination (e.g. from the nearby input coupler).

The $1/9\,\pi$ mode which has the least field in the end cells exhibits a residual resistance down to $1.4\,\nano\ohm$ in the best case. The $\pi$ mode with equal field distribution in each cell shows an intermediate \rres of $6.4\,\nano\ohm$ while the $8/9\,\pi$ mode with maximum field in the end cells has the highest \rres of $9.9\,\nano\ohm$. If the low level of magnetic field in the center of the shielding would extend to the end cells, the low surface resistance, corresponding to a quality factor of more than $10^{11}$ at $1.5\,\kelvin$, might be achieved in the $\pi$ mode as well.

In conclusion we can say that we observed a significant contribution to the residual resistance that increases with the temperature difference between the ends of the cavity tank system that is installed in a horizontal test stand. Given a high-Q cavity with an overall surface resistance of order $10\,\nano\ohm$ at $2\,\kelvin$ this effect could dominate the losses.

\section{Simulation of thermoelectric induced magnetic flux\label{sec:sims}}

\subsection{Setup for simulations: Symmetry and symmetry breaking \label{sec:sym}}
\begin{figure}
\includegraphics[width=\linewidth]{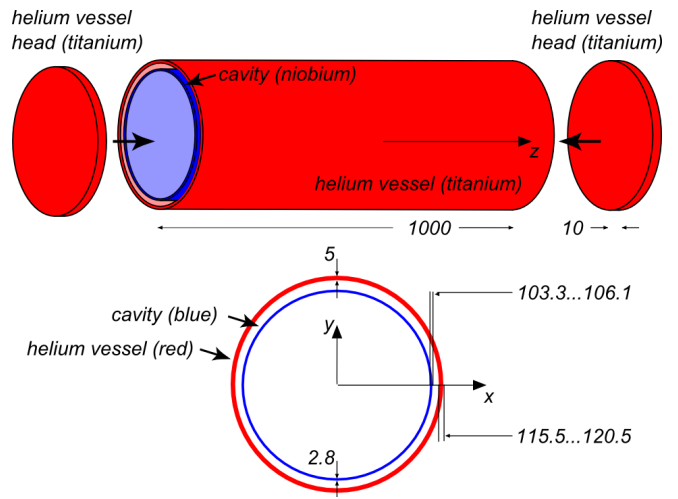}
\caption{Simplified geometry for COMSOL simulations. Dimensions are in millimeters.\label{fig:COMSOLsetup}}
\end{figure}
To investigate the experimental results further and to make a quantitative analysis of the thermocurrents we turned to numerical simulations of the system. In particular it is important to understand the distribution of the currents and the magnetic flux. In our previous paper, we made a rough estimate of the induced current and the corresponding magnetic flux. Now, we want to use temperature dependent material properties to yield a more reliable number on the amount of flux which is to be expected on the RF surface. For that purpose we conducted COMSOL simulations which will we discuss in the next section.

We modeled a simplified geometry consisting of two coaxial cylinders (Figure \ref{fig:COMSOLsetup}). The inner cylinder which resembled the cavity had the typical cavity wall thickness of $2.8\,\milli\meter$ and was defined as niobium. The outer cylinder and the two end plates resembled the helium vessel and helium vessel heads and were defined as titanium with a thickness of $5\,\milli\meter$ (vessel) or $10\,\milli\meter$ (vessel heads) respectively. The $1\,\meter$ length of both cylinders was the approximate length of a TESLA cavity in a tank. The model exhibited perfect cylindrical symmetry.

\begin{figure}
\includegraphics[width=.8\linewidth]{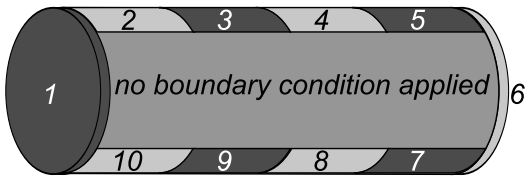}
\caption{Different areas in which thermal boundaries could be applied in COMSOL simulations. No boundary condition was applied to the rest of the setup.\label{fig:boundaries}}
\end{figure}

The simulations were performed in several steps. For each step different temperature boundary conditions were applied. We defined ten areas on the outer surface of the tank (Figure \ref{fig:boundaries}). Each area could either be assigned a fixed temperature for the simulation or left undefined as the rest of the setup. Thereby we created temperature distribution similar to our experimental situation.

In the first step, we defined the temperatures only on the end plates, areas 1 ($10\,\kelvin$) and 6 ($100\,\kelvin$). In this case the isothermals were perpendicular to the cylinder axis. The resulting current distribution was cylindrically symmetric and the azimuthal magnetic flux on the RF surface was zero though it increased rapidly as one moves into the wall. This situation is as expected from Gauss' Law.

In the next steps, the areas 2 -- 5 and 7 -- 10 were assigned fixed temperatures as well, leading to asymmetric temperature distributions. The isothermals were not orthogonal to the cylinder axis anymore but distorted. Thereby the electrical symmetry was broken due to the temperature dependence of the electrical resistance. This resembles the cavity cooldown when liquid helium started to collect first at the bottom of the tank.

Simulations where the inner cylinder is replaced by the actual cavity geometry are not presented in this paper but are in preparation.

\subsection{Input data} \label{sec:thermopower}
For these simulations the temperature dependent material properties are needed. We used literature data as given in references \cite{Padamsee2009,Jensen1980,Merio2011} for the temperature dependent electrical conductivity. The thermal conductivity was set constant to improve the convergence of the COMSOL simulations. We choose $\kappa\sub{Nb}\,=\,50\frac{\watt}{\meter \cdot \kelvin}$ and $\kappa\sub{Ti}\,=\,22\frac{\watt}{\meter \cdot \kelvin}$ as approximations \cite{Kes1974,Jensen1980,Merio2011}.

On the thermopower only poor information is available. Hence we performed sample measurements of the thermopower on niobium ($RRR\,=\,300$) and titanium (grade 2). The dimension of the samples were approximately $2\,\milli\meter \times 5\,\milli\meter \times 30\,\milli\meter$ and $2\,\milli\meter \times 2.3\,\milli\meter \times 30\,\milli\meter$. 

The experiments were performed with a physical property measurement system by Quantum Design \footnote{\url{www.qdusa.com/sitedocs/productBrochures/1070-002.pdf}} using the thermal transport option \footnote{\url{http://www.qdusa.com/sitedocs/productBrochures/tto_rev7-06.pdf}}. 

The system allowed high precision measurements of the thermopower over a wide temperature range. We measured all four samples (two niobium, two titanium) in the range from $4.5\,\kelvin$ to $300\,\kelvin$ several times. 

\begin{figure}
\includegraphics[width=.95\linewidth]{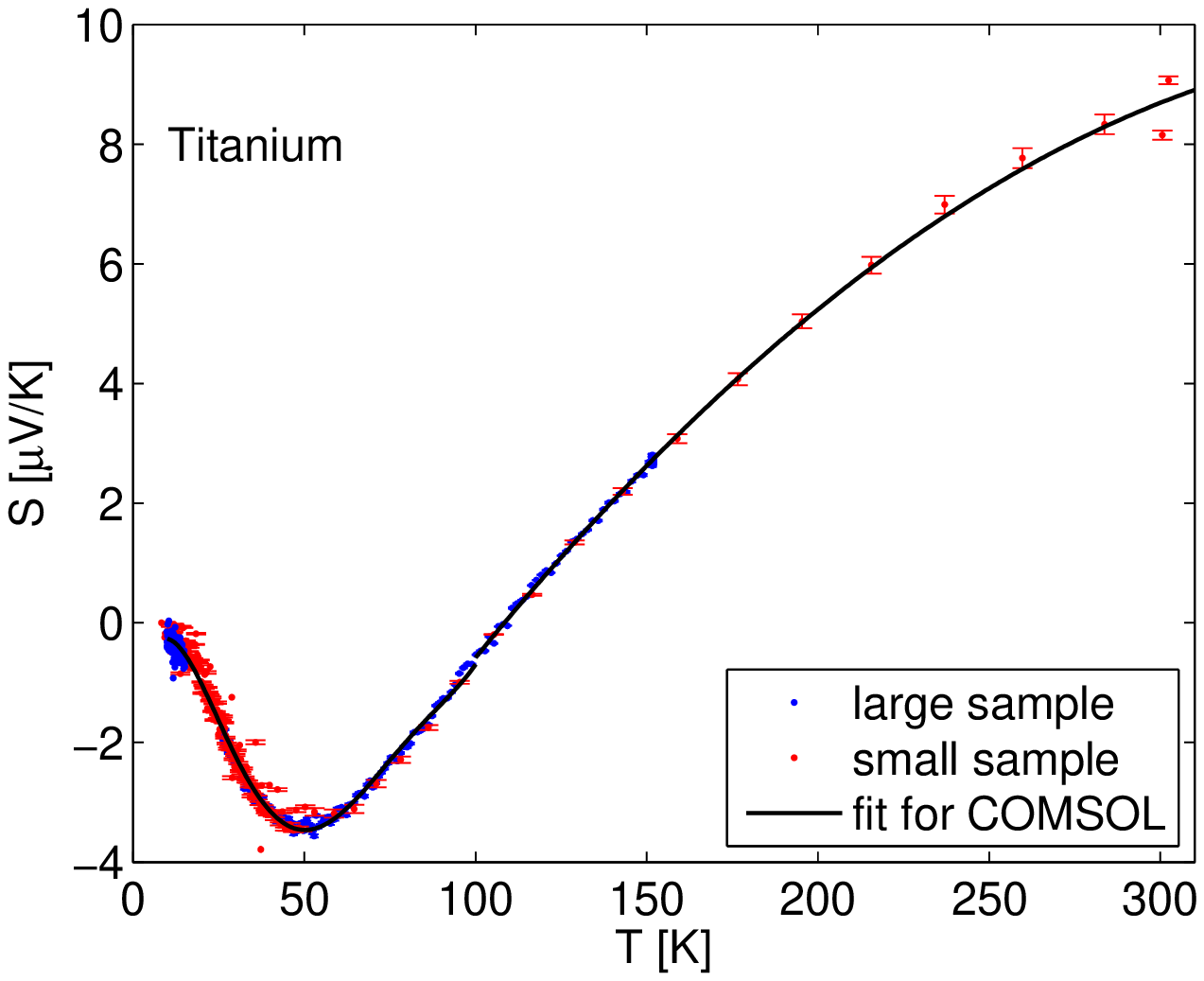}
\includegraphics[width=.95\linewidth]{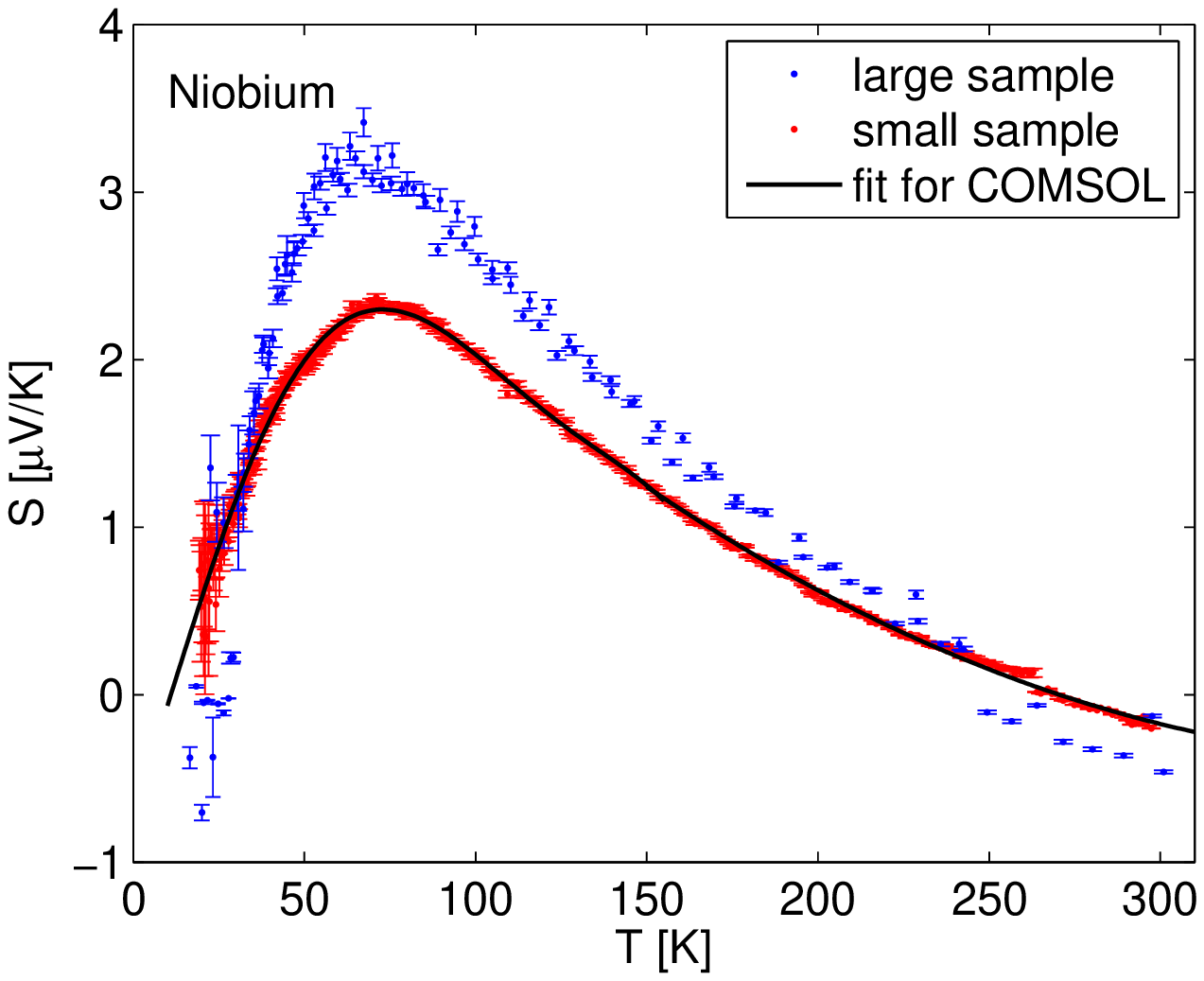}
\caption{PPMS thermopower data of niobium and titanium for the small ($2\,\milli\meter \times 2.3\,\milli\meter \times 30\,\milli\meter$) and the large sample ($2\,\milli\meter \times 5\,\milli\meter \times 30\,\milli\meter$). \label{fig:S}}
\end{figure}

Figure \ref{fig:S} shows the measured thermopower $S\,[\micro\volt/\kelvin]$ as a function of temperature. The data is in agreement with literature data \cite{Blatt1976,Nystrom1959,Weinberg1966}. For the titanium data, the small and the large sample were the same. The curve was fitted as shown in Figure \ref{fig:S}. A table was extracted and entered into COMSOL.

\begin{figure*}
\includegraphics[width=\linewidth]{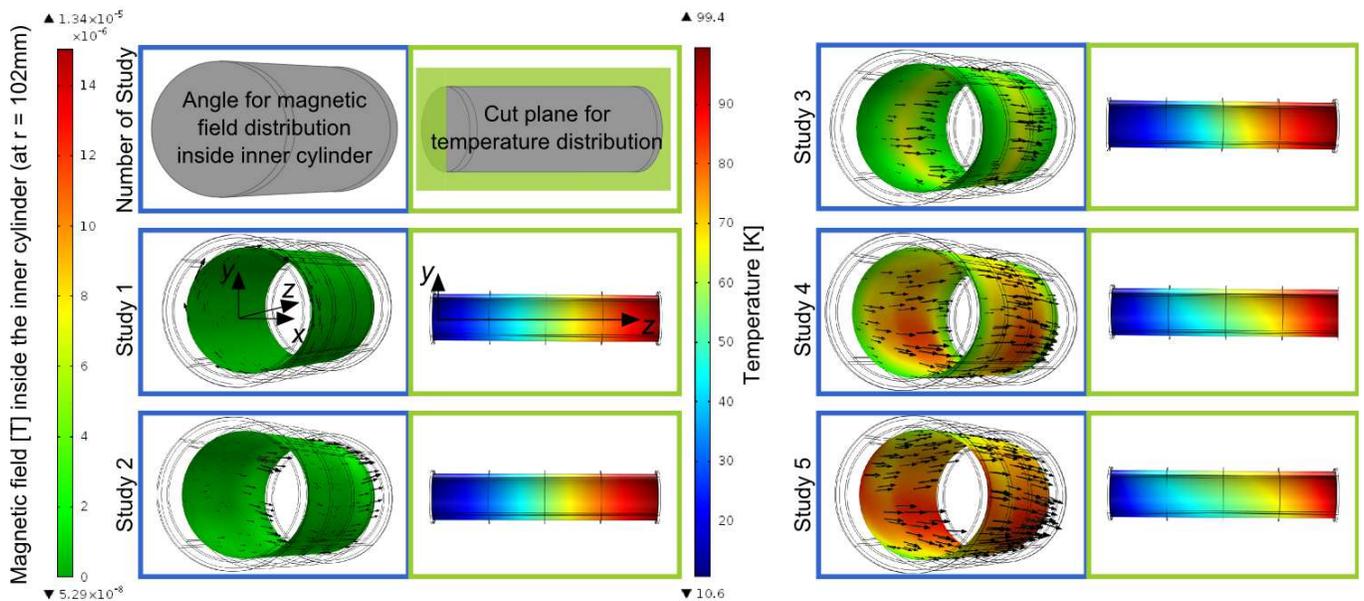}
\caption{Results of the COMSOL simulations. The left cylinders display the magnetic field distribution inside the cylinders at $r\,=\,102\,\milli\meter$. The legend is given on the left reaching from $\approx 50\,\nano\tesla$ (green) to $\approx 13\,\micro\tesla$ (red). The right figures display a cut through the temperature distribution inside the cylinders. The legend for the temperature reaches from $10\,\kelvin$ to $100\,\kelvin$.  \label{fig:SimResB}}
\end{figure*}

For niobium, the two samples did not give the same result. In general, both curves are similar and have the same zero-crossing but the amplitude was higher for the large sample compared to the small sample. The COMSOL simulations were first performed with the fitted data for the small sample as shown in Figure \ref{fig:S} to obtain a conservative lower limit. Note that the disagreement in the measured thermopower in the temperature range up to $75\,\kelvin$ is only in the $30\,\%$ range. The simulations were also repeated with the less conservative Seebeck values for a sensitivity check, but we found that the impact was not dramatic (see later Figure \ref{fig:Bsim}). Nevertheless we intend to repeat the measurements in the future to provide more accurate results.

\subsection{Magnetic flux at the RF surface} \label{sec:simres}

We performed five simulations:

\begin{enumerate}
	\item A symmetric test distribution as a consistency check for the simulations. Surface 1 was set to $10\,\kelvin$ and surface 6 to $100\,\kelvin$.
	\item Asymmetric studies: Surfaces 1 and 10 at $10\,\kelvin$ and surfaces 5 and 6 at $100\,\kelvin$
	\item Surfaces 1, 10 and 9 at $10\,\kelvin$ and surfaces 4, 5 and 6 at $100\,\kelvin$
	\item Surfaces 1, 10, 9 and 8 st $10\,\kelvin$ and surfaces 3, 4, 5 and 6 at $100\,\kelvin$
	\item Surfaces 1, 10, 9, 8 and 7 at $10\,\kelvin$ and surfaces 2, 3, 4, 5 and 6 at $100\,\kelvin$
\end{enumerate}

The simulations were performed as two steps. First, the temperature distribution and the resulting electric current distribution was computed. In the second study, this input data was used to calculate the magnetic field distribution.

The RF surface is at a radius of $r\,=\,103.3\,\milli\meter$. Figure \ref{fig:SimResB} shows the solution for the magnetic field \textit{inside} the cylinders close the RF surface at $r\,=\,102\,\milli\meter$ to ensure that it is inside the cylinder and not in the cylinder wall due to meshing. The arrows display the direction of the magnetic field at the same radius. The arrow size is proportional to the field strength. Generally the field is near perpendicular to the surface in high field regions. 

In addition to the magnetic field distribution, Figure \ref{fig:SimResB} displays cuts with the temperature distributions inside the cylinders.

\begin{figure}
\includegraphics[width=\linewidth]{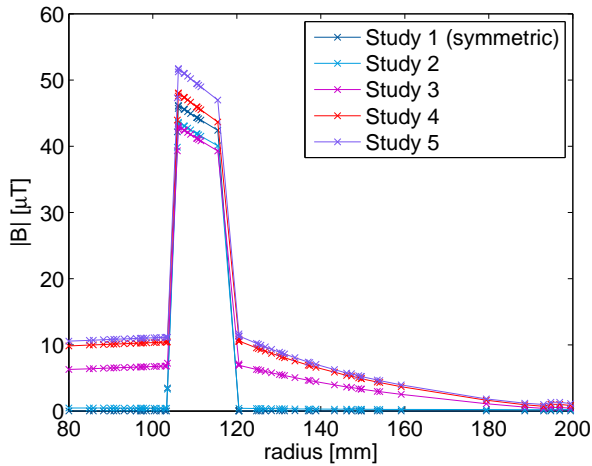}
\caption{Absolute magnetic field along a cut trough the cylinders at $z\,=\,0.5\,\meter$. \label{fig:SimBCut}}
\end{figure}

Figure \ref{fig:SimBCut} displays the magnetic field at a cut through the walls of the two cylinders at $z\,=\,0.5\,\meter$. For the symmetric case, no field is inside the inner cylinder, the field increases in the inner cylinder wall and decreases towards the outer cylinder. Outside of the two cylinders there is no magnetic field anymore. This is just as expected from Gauss' Law. For studies 2 to 5, the increasing asymmetry of the current density leads to an increased magnetic field inside the cylinders. This field can get trapped during superconducting phase transition and produce RF losses as we have observed in the experiment.

\begin{figure}
\includegraphics[width=\linewidth]{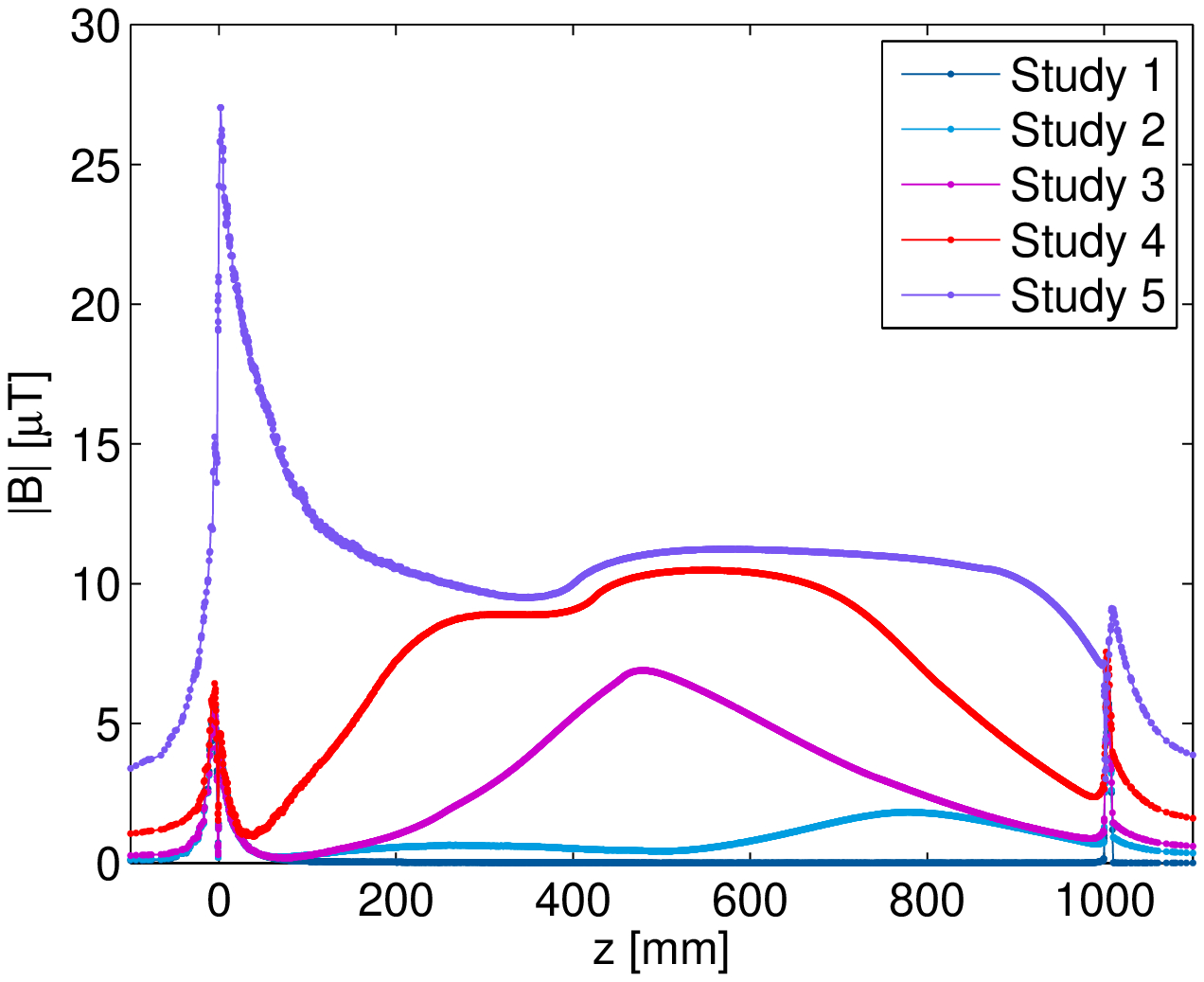}
\includegraphics[width=\linewidth]{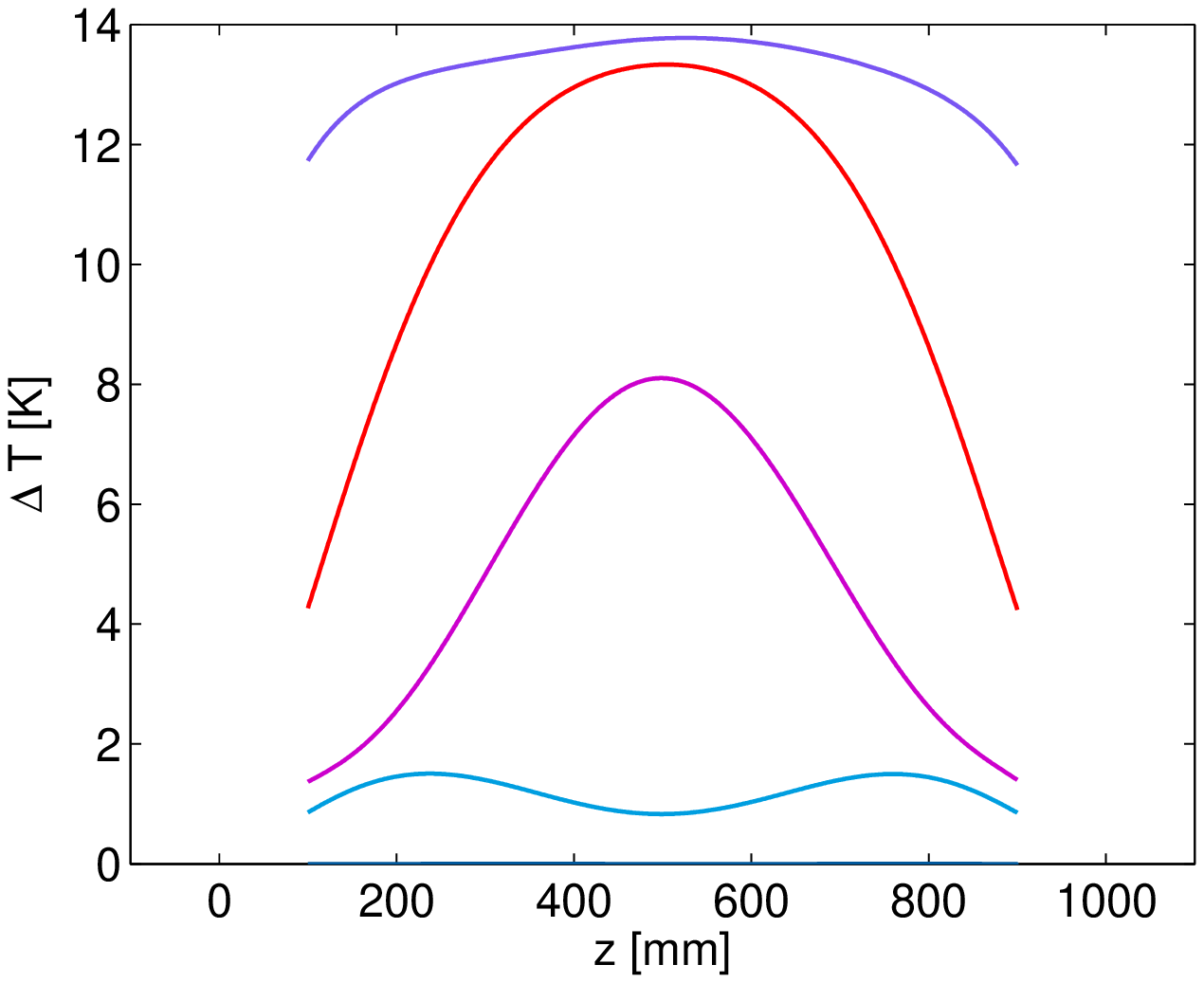}
\caption{Top Figure: Absolute magnetic field along $z$ at $x\,=\,102\milli\meter$ and $y\,=\,0\,\milli\meter$. The peaks at $z\,=\,0\,\milli\meter$ and $z\,=\,1\,\meter$ are due to the sharp corners when the cylinder passes into the end plates. Bottom Figure: Temperature difference between top and bottom of the inner cylinder.  \label{fig:SimZ}}
\end{figure}

Figure \ref{fig:SimZ} displays a cut of the RF surface along the z-axis. The position of the cut was chosen to be at $x\,=\,102\,\milli\meter$, $y\,=\,0\,\milli\meter$. Furthermore the Figure displays the temperature difference between top and bottom of the inner cylinder. Where the temperature difference is maximum, the inhomogeneity in the electric current distribution is largest and hence the most magnetic field penetrates the RF surface.

The results clearly illustrate that a non symmetric temperature distribution leads to a nonzero magnetic field on the RF surface. The highest fields are obtained where the temperature difference between top and bottom part are largest. In study 5, this effect leads to a more or less constant magnetic flux along the cylinder while in study 3 where the temperature gradient from bottom to top is less distinct the magnetic field increases towards the center of the cylinders ($z\,=\,0.5\,\meter$). In the experiment, the highest degree of broken symmetry is obtained when the first superconducting path through the cavity establishes below $9.2\,\kelvin$. This corresponds to an extreme example of case 5 where the DC resistance of the bottom part of the inner cylinder drops to zero while the rest of the cavity is still normal conducting and hence the thermocurrent is forced into the sc path. In this case all cells would be affected by the additional induced flux in the same way which is in agreement with the linear increase in Figure \ref{fig:ModeCompare} with slope $\approx\,1$. Temperature mapping data of a dressed multicell cavity in a horizontal test would give more insight in the real temperature distribution.

\begin{figure}
\includegraphics[width=\linewidth]{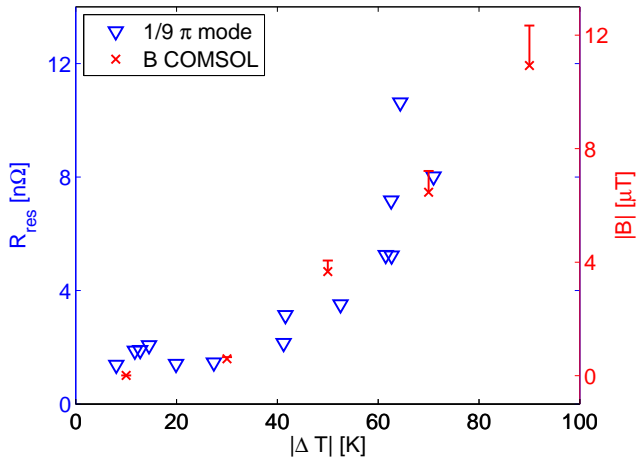}
\caption{Magnetic field at the RF surface simulated for different temperature differences in combination with the data of Figure \ref{fig:RvsDT}. The red bars indicate the magnetic field that is obtained assuming a $30\%$ higher thermopower, corresponding roughly to the uncertainty resulting from the discrepancy depicted in Figure \ref{fig:S}. \label{fig:Bsim}}
\end{figure}

Finally, we varied the temperature difference in the simulation and set the warmer boundaries of study 5 to the temperatures $80\,\kelvin$, $60\,\kelvin$, $40\,\kelvin$ and $20\,\kelvin$ while the other surfaces remained fixed at $10\,\kelvin$. The result is given in Figure \ref{fig:Bsim} together with the experimental findings. Since the measurements of the thermopower of niobium allowed for $30\,\%$ higher values, we added bars indicating the magnetic field at the RF surface for a $30\,\%$ higher thermopower of niobium. The results show that the magnetic field increases with the temperature difference. While we do not claim that the quantitative values should be directly compared, the trend is very similar. 

In conclusion, we assess that the COMSOL simulations support the assumption that thermoelectrically induced magnetic field penetrates the RF surface due to an inhomogeneous temperature distribution. We see that even for a moderate asymmetry the flux is locally of order several $\micro\tesla$. The RF losses resulting from the increased magnetic field will depend on the fraction of the nine cell cavity that is affected by the thermocurrent due to more pronounced asymmetry, the exact location of the induced magnetic field relative to the RF field and the amount of flux that eventually gets trapped in the material. However the variation of about $10\,\nano\ohm$ seen in the cavity tests are consistent with the thermocurrent hypothesis.

\section{Summary and Outlook\label{Chap:Sum}}

We demonstrated that poorly controlled cooling conditions can significantly deteriorate the quality factor of SRF cavities due to the generation of thermocurrents. While already a simple misalignment of the cavity to the helium tank could break the symmetry of the cavity tank system, although the effect on \rs should be small \cite{Crawford2013} even a mechanically cylindrically symmetric system will generate magnetic flux on the RF surface. The presented simulations explain how the electrical symmetry is broken due to the temperature dependance of the electrical conductivity.  

We also demonstrated that an initial cooldown through \tc with small $\Delta\,T$ can in principle yield residual resistances at $1\,\nano\ohm$ (parked cooldown), \textit{provided the external magnetic shielding is very effective in eliminating external residual flux}.

What we did not investigate in this paper is the influence of changed cooling conditions on the efficacy of the Meissner effect as we explained in section \ref{sec:trapp} since our magnetic shield eliminated most of the residual field. 

Temperature mapping during dressed cavity horizontal test could improve the understanding on the real temperature distribution during cooldown and cycles. Furthermore it would help with characterization of cooling parameters like cooling speed, temperature gradients in all spatial directions and homogeneity.

\begin{acknowledgments}
We would like to thank Dr.~Axel Neumann and Dr.~Adolfo Velez for providing the simulations of the field distribution of the TESLA cavity and the accompanying bead pull measurement. We thank Michael Schuster, Andre Frahm, Sascha Klauke, Dirk Pfl{\"u}ckhahn, Stefan Rotterdam and Axel Hellwig for experimental support. Furthermore we thank the group of Dr. Klaus Kiefer for the support with the thermal transport measurements, Bernd Spaniol (Hereaus) for providing the niobium and our machine shop for support especially with last minute modification of the samples.
\end{acknowledgments}

\bibliographystyle{apsrev4-1}
\bibliography{lit}

\end{document}